\documentclass[useAMS]{mn2e}
\usepackage{graphicx}
\usepackage{amsmath}
\usepackage{amssymb}
\usepackage{color}
\usepackage{url}

\newcommand\nodata{ ~$\cdots$~ }%

\newcommand  \acc     {\ifmmode {\rm km\,s}^{-2} \else km\,s$^{-2}$\fi}

\newcommand  \ergs     {\ifmmode {\rm ergs\,s}^{-1} \else ergs s$^{-1}$\fi}
\newcommand  \ergcms   {\ifmmode {\rm erg~cm}^{-2}\,{\rm s}^{-1}
                        \else erg~cm$^{-2}$\,s$^{-1}$\fi}
\newcommand  \ergcmsA  {\ifmmode{\rm erg\,cm}^{-2}\,{\rm s}^{-1}\,{\rm\AA}^{-1}
                        \else ergs\,cm$^{-2}$\,s$^{-1}$\,\AA$^{-1}$\fi}
\newcommand  \ergcmsHz {\ifmmode{\rm ergs\,cm}^{-2}\,{\rm s}^{-1}\,{\rm Hz}^{-1}
                        \else ergs\,cm$^{-2}$\,s$^{-1}$\,Hz$^{-1}$\fi}
\newcommand  \phcms    {\ifmmode {\rm ph\,cm}^{-2}\,{\rm s}^{-1}
                        \else ph\,cm$^{-2}$\,s$^{-1}$\fi}
\newcommand  \phcmsA   {\ifmmode {\rm ph\,cm}^{-2}\,{\rm s}^{-1}\,{\rm\AA}^{-1}
                        \else ph\,cm$^{-2}$\,s$^{-1}$\,\AA$^{-1}$\fi}
\newcommand \msun {M$_\odot$}
\newcommand\aj{{AJ}}% 
\newcommand\apj{{ApJ}}% 
\newcommand\apjl{{ApJ}}% 
\newcommand\apjs{{ApJS}}% 
\newcommand\aap{{A\&A}}% 
\newcommand\mnras{{MNRAS}}% 
\newcommand\pasj{{PASJ}}% 
\newcommand\nat{{Nature}}% 
\title[SDSS2 Supernova Delay Time Distribution]
{
The Delay Time Distribution of Type-Ia Supernovae from Sloan II}
\author[D. Maoz, F. Mannucci, and T.D. Brandt]
{Dan Maoz$^{1}$\thanks{E-mail: maoz@astro.tau.ac.il},
Filippo Mannucci$^{2}$, Timothy D. Brandt$^{3}$\\
$^{1}$School of Physics and Astronomy, 
Tel-Aviv University, Tel-Aviv 69978,
Israel\\
$^{2}$INAF - Osservatorio Astrofisico di Arcetri, Largo Enrico Fermi 5, Firenze
50125, Italy\\
$^{3}$Department of Astrophysical Sciences, Ivy Lane, Princeton University,
Princeton, NJ 08540, USA\\
} \date{\today}

\begin{document}

\maketitle

\label{firstpage}

\begin{abstract}
We  derive
the delay-time distribution (DTD) of type-Ia supernovae (SNe Ia)
using
a sample of 132 SNe Ia, discovered by the Sloan Digital Sky Survey II (SDSS2) 
among 66,000 galaxies with spectral-based star-formation histories (SFHs). 
To
recover the best-fit DTD, the SFH of
every individual galaxy is compared, using  Poisson statistics, 
to the number of SNe that it hosted (zero or one), based on 
the method introduced in Maoz et al. (2011).
 This SN sample
differs from the SDSS2 SN Ia sample analyzed by Brandt et
al. (2010), using a related, but different, DTD recovery method.
Furthermore, we use a simulation-based SN detection-efficiency function, 
 and we apply a number of important corrections to the galaxy SFHs and
 SN~Ia visibility times. 
%Using the core-collapse SNe in SDSS2, we
%show that this sample, with its higher redshift compared to the Lick 
%Observatory SN Search sample analyzed by Maoz et al., is much less
%affected by the limited SDSS fibre aperture, which caused a ``leak''
%of DTD power to late time bins. 
The DTD that we find has $4\sigma$ detections
in all three of its time bins: prompt ($\tau<0.42$~Gyr),  intermediate
($0.42<\tau<2.4$~Gyr),  
and delayed ($\tau>2.4$~Gyr), indicating a continuous DTD, and it is among the most accurate and
precise among recent DTD reconstructions. The best-fit power-law form to the recovered DTD is
$t^{-1.12\pm0.08}$, consistent with generic $\sim t^{-1}$ predictions of SN Ia progenitor
models based on the gravitational-wave-induced mergers of binary white dwarfs.
The time integrated number of SNe Ia per formed stellar mass 
is $N_{\rm SN}/M=0.00130\pm 0.00015$~\msun$^{-1}$,  or about 4\% of the
stars formed with initial masses in the $3-8$~\msun\ range. This is lower than,
but largely consistent with, 
several recent DTD estimates based on SN rates in galaxy clusters and in
local-volume galaxies, and is higher than, but consistent with $N_{\rm SN}/M$
estimated by comparing volumetric SN Ia rates to cosmic SFH.
  
\end{abstract}


%.

%\newpage

\newpage

\section{Introduction}

What exactly it is that explodes in 
Type-Ia supernovae (SNe Ia) is unknown -- this is the SN Ia progenitor problem
 (see Howell 2011; Maoz \& Mannucci 2012, for recent reviews). Based on
the energetics and chemical makeup of the ejecta, the explosion involves
the thermonuclear combustion of a
carbon-oxygen white dwarf (WD) into iron-peak elements. However, the
identity of the exploding system, and how it ignites, are unknown. 
The repercussions of this problem
include
the risk of systematic errors in cosmological inferences using
SNe Ia as distance indicators, an incomplete picture of cosmic
history (we do not know the physical identity of one of the main metal
producers), and ambiguity regarding the final outcome of two important stages in
stellar and binary evolution -- cataclysmic binaries and WD
mergers.

Two main
progenitor scenarios are usually considered.
In the single degenerate (SD)
model (Whelan \& Iben 1974; Nomoto 1982), a carbon-oxygen
white dwarf (WD) grows in mass through accretion from
a non-degenerate stellar companion -- a main sequence star, a subgiant,
a helium star (i.e. a stripped evolved star),
or a red giant -- until it approaches the
Chandrasekhar mass, ignites, and explodes in a thermonuclear runaway.
In the double degenerate (DD) scenario (Webbink 1984; Iben \& Tutukov 1984),
two WDs merge after losing energy and
angular momentum to gravitational waves, with
the more massive WD tidally disrupting and accreting
the lower-mass object, and at some point igniting.

To date, neither the SD nor the DD model is clearly favored
observationally.
Both models, SD and DD, suffer from problems,
theoretical and observational. One example is
the lack of early-time UV shock signatures (Kasen 2010)
due to ejecta hitting
a SD red-giant donor (e.g., Nugent et al. 2011; Brown et al. 2012; 
Bloom et al. 2012),
or lack of early radio and X-ray
emission,
expected if the SN
blast wave were to encounter   a circumstellar wind from a SD donor
(e.g. Chomiuk et al. 2012,
Horesh et al. 2012).
Perhaps most dramatically, Schaefer \& Pagnotta (2012) have shown
that there is no remaining companion, down to 
luminosity limits corresponding to
main-sequence K stars, in the remnant of a SN Ia from {\it circa} 1600
(confirmed as such with a light-echo spectrum by Rest et al. 2008).
These observations disfavor non-degenerate donors.

On the other hand, calcium, sodium, and other
absorption lines possibly associated with a SD donor star have been
seen in some SN Ia spectra (Patat et al. 2007; Simon et al. 2009; 
Sternberg et al. 2012; Dilday et al. 2012). 
Against the DD model, it has also long
been argued that the merger of two unequal-mass WDs will lead to
an accretion-induced collapse to a neutron star,
i.e. a core-collapse SN (although not a typical one, due to the
absence of an envelope), rather than a SN Ia (Nomoto \& Iben 1985;
Shen et al. 2012), but ways
have been proposed to avoid this
(e.g. Pakmor et al. 2010;
Van Kerkwijk et al. 2010).
Several ``SN on hold''
scenarios have also been proposed (Di Stefano et al. 2011; Justham
2011; Kashi \& Soker 2011; Ilkov \& Soker 2012), where
ignition is potentially long delayed. During
this time the traces of
the messy accretion process (or even of the donor itself) could
disappear. More than one SN Ia
channel could be at work, and this possibility has often been raised
(e.g. Dilday et al. 2012).

It has long been realized that SN Ia rates
may discriminate among progenitor
models, via the ``delay time distribution'' (DTD).
The DTD is the hypothetical SN
rate versus time that would follow a brief burst of star formation.
 It is the
``impulse response'' that embodies the
physical information of the system, free of nuisances such as, in the present
context, the star-formation histories (SFHs)
of the galaxies hosting the SNe.
It is directly linked to the stellar  and
binary evolution timescales up to the
explosion, so different progenitor scenarios predict
different DTDs. Furthermore, apart from the progenitor problem, the SN DTD,
together with SFH, dictates the heavy-element chemical evolution of
individual galaxies and of the Universe as a whole. A 
well-determined DTD based on observations would be a valuable input to
metal-enrichment history.
 Various theoretical DTDs
have been proposed, whether based on detailed
binary population synthesis (e.g., Yungelson \& Livio 2000; Nelemans
et al. 2001; Han \& Podsiadlowski 2004; Mennekens et al. 2010; 
Ruiter et al. 2011; Meng et al. 2011; see Wang \& Han 2012, for a review),
or on some simple physical considerations.
In the DD model, the event rate ultimately depends on the loss of energy
to gravitational radiation. If, at the end of the last common-envelope
phase, the WD separation distribution follows a power law, then for a
fairly large range of power-law indices around $-1$, the   
DTD for this  model will also be a power-law in 
time, out to a Hubble time, $\sim t^{\beta}$, with $\beta$ not far from
$-1$. Of course, the post-common-envelope separation distribution
could be radically different, and not even be well-approximated by 
a power law, and thus a $\sim t^{-1}$ DTD form should not be
considered unavoidable (see, e.g., Ruiter et al. 2011). Nevertheless, the
 WD separation distribution that emerges from simulations does seem
close to a power law of index $-1$ (see discussion in Maoz et al. 2012).   
In the SD models, in contrast, the DTD often
cuts off after a few Gyr, due to a lack of evolved donors that can
transfer enough mass to the WD. Here again, this expectation could be 
modified in the ``on hold'' scenarios mentioned above, which could
introduce additional delays before explosion.

Mannucci et al. (2005, 2006) and Sullivan et al. (2006), 
analyzing  SN Ia rates as a function of galaxy colors,
made the first direct estimates of SN Ia delay times, finding
evidence for a range of delay times, with 
both ``prompt'' SNe Ia that explode within hundreds
of Myr of star formation, and ``delayed'' SNe Ia that explode in populations
of ages of at least a few Gyr. Totani et al. (2008) measured SN Ia rates
in elliptical galaxies as a function of luminosity-weighted galaxy age, 
modeled by assuming a coeval stellar population, 
and deduced a DTD consistent with a $t^{-1}$ form.
Recently, a number of novel DTD recovery techniques,
using a variety of  SN samples, environments, and redshifts,
have been yielding mostly
 consistent DTDs.
One approach to DTD recovery has been to measure
SN Ia rates versus redshift in galaxy clusters, in which the
observed SN Ia rate vs. cosmic time since the stellar formation epoch
provides an almost direct measurement of the DTD. 
 Another approach has been to measure the volumetric SN rate vs. redshift from field
surveys. This rate will be the convolution of the DTD with the cosmic 
SFH. The cluster-based SN~Ia DTD, as analyzed in
Maoz et al. (2010), as well as the one emerging from volumetric SN Ia rates
(Graur et al. 2011), also  appear to follow the $\sim t^{-1}$ form
generically expected in the DD picture, although some controversies remain
(see Maoz \& Mannucci 2012). Very recent new and precise volumetric
rates out to redshift $z=1$
by Perrett et al. (2012) merge smoothly with the Graur et al. (2011)
rates at higher $z$, and their analysis confirms the above DTD result.

The above approaches to DTD recovery involve loss
of information, by associating all of the SNe found at a given
redshift with all the galaxies at that redshift.
Brandt et al. (2010; B10) and Maoz et al. (2011; M11) simultaneously 
introduced two closely related new methods for DTD reconstruction
that avoid this averaging. 
Rather than using SFHs averaged over  many galaxies in some
redshift bin, and comparing to the number of SNe found by a survey at
those
redshifts, the methods take into account the detailed SFH of each
individual galaxy monitored by a SN survey.
In the method used by M11, the expected
SN rate, for some assumed DTD, is compared 
to the observed SN rate in each galaxy.
The DTD parameter space is searched for the DTD that best reproduces 
the observed individual galaxy SN rates. M11 applied the method
to a subsample of the galaxies in the Lick Observatory SN Search
(LOSS, Leaman et al. 2011, Li et al. 2011a,b) that have SFHs based on 
Sloan Digital Sky Survey (SDSS; York et al. 2000)
 spectra, as recovered by Tojeiro et
al. (2009), and to the SNe found by LOSS in those galaxies. 
Using a DTD with three time bins, 
prompt ($t<0.42$~Gyr), ``intermediate'' 
($0.42<t<2.4$~Gyr), and delayed ($t>2.4$~Gyr),
M11 found significant detections
of both prompt and delayed SNe Ia, and an 
overall DTD consistent with a $t^{-1}$ form.   

The DTD recovery method of B10 is similar to that of M11
in several respects. However,
in B10, rather than (as in M11) comparing the SN 
numbers observed in each galaxy to the predictions of DTD models, 
the mean spectrum of the SN host galaxies is compared to mean spectra
formed with mock host-galaxy samples that result from
each of the tested DTD models. B10 applied their method
to a sample of 77,000 galaxies in the ``Stripe 82'' region of the SDSS, 
having spectra with Tojeiro et al. (2009) SFH reconstructions, and to the 
SNe Ia that they hosted in the SDSS-II SN survey (SDSS2; 
Frieman et al. 2008; Sako et al. 2008).
Like M11, B10
detected, among their three SN Ia DTD time bins, significant signals 
in the $t<420$~Myr and $t>2.4$~Gyr bins. Furthermore, B10 separated
their SNe Ia, based on light-curve shape,
 into  ``low-stretch'' and ``high-stretch'' subsamples,
and found that 
high-stretch SNe have a DTD signal mainly in the prompt bin, while
low-stretch SNe have a DTD dominated by the delayed bin. 

In Maoz \&  Badenes
(2010), the M11 method was applied to a sample of SN remnants in the
Magellanic Clouds, treating them as an effective SN survey, one in which
the events are visible for tens of kyr. 
The SN rates in individual regions of the
Clouds were compared to the detailed 
SFH of each region, as reconstructed by Harris \& Zaritsky (2004, 2009)
using isochrone fitting to
the resolved stellar populations. In this
sample, 
the limited number of SN remnants that originated from SNe Ia, of
order 10, 
limited the precision of the DTD, and hence
allowed the significant detection only of the prompt SN Ia component.
 
In the present paper, we apply the M11 DTD recovery method to a sample of SNe
discovered in SDSS2. 
There are several differences in our analysis compared to the
B10 analysis of the SDSS2 SNe. The DTD recovery approach, as
described above, is different. We show, however, that the M11 method, when 
applied to the same sample used by B10, and 
using the same assumptions as B10,
does give 
a DTD similar to the one found by B10, indicating the approaches 
are equivalent.
We construct a new SDSS2 SN Ia sample with careful attention to selection
criteria, SN classification, and SN-host matching, resulting in a 
sample that is composed of SN events that are largely distinct from those in
the B10 sample. With 132 SNe Ia, our sample
is somewhat larger than the B10 sample, with its 101 SNe,
and it is more complete, which is important in the DTD context. 
 Finally, in our
analysis, we use an empirical detection efficiency function from the
SDSS2 project, and we apply several necessary corrections, not included in B10,
 to the calculation 
of the formed stellar mass in the monitored galaxies, and to the 
SN visibility time of each galaxy. These corrections do not affect
much the form of the DTD, but do affect its normalization, resulting 
in a more robust time-integrated DTD. Further details on each of these
new aspects of the present work are given in the body of this paper.

\section{Data}
\label{datasection}
\subsection{Galaxy sample}
\label{galaxysample}
The SDSS (York et al. 2000), in its first stage, 
was a survey of $\sim 10^4~{\rm deg}^2$ of the north
Galactic cap, consisting of imaging 
in five photometric bands ($u, g, r, i, z$), and $3''$-aperture  fibre
spectroscopy of $\sim 10^6$ targets, mostly galaxies, with $r\lesssim 18$~mag. 
Tojeiro et al. (2009) performed spectral synthesis modeling of 
all galaxy spectra in the SDSS using their VESPA code (Tojeiro et al. 2007).  
VESPA uses all of the available absorption
features, as well as the shape of the continuum,
to deconvolve the observed spectra and obtain an 
estimate of the SFH. The SFH of each galaxy
consists of the amounts of stellar mass formed in fixed bins of
look-back time. 
In order to recover the maximum amount of reliable information,
the number of time bins used 
is variable, and depends on the quality of the data
on each galaxy. At the highest resolution, VESPA uses
16 age bins, logarithmically spaced between
0.002 Gyr and $t_0=13.7$~Gyr, the age of the Universe. 
When data do not have sufficiently high signal-to-noise ratio
for a fully resolved reconstruction, 
pairs of adjacent time bins are averaged.
This process may be repeated down to the last two remaining bins.
In the end, VESPA recovers each galaxy's SFH using a different 
set of time bins.

VESPA masses are calculated 
assuming a Kroupa (2007) initial-mass function (IMF); all of our results
will therefore include this assumption implicitly.
Bell et al. (2003) have shown that the Kroupa IMF gives a
similar total stellar mass to that of a ``diet Salpeter''  IMF,
obtained by multiplying by 0.7 the total mass of the original 
Salpeter (1955) IMF, to account for the reduced number of low-mass
stars. Thus, our results will be comparable to other SN rate studies,
such as Mannucci et al. (2005, 2006), and other recent DTD reconstructions (Maoz et
al. 2010; Maoz \& Badenes 2010; M11; Graur et al. 2011) 
that have assumed the diet-Salpeter IMF.

During 3 years from 2005 to 2007, for 90 days a year, 
the SDSS2 repeatedly imaged, with a 4-day cadence, Stripe 82, 
 a $120^{\circ}\times 2.5^{\circ}$ equatorial region,
in the RA range from $-60^\circ$ to $+60^\circ$
(Frieman et al. 2008; Sako et al. 2008). As
described in B10, in Data Release 7 (DR7) of the SDSS there are about
77,000 Stripe-82 non-active galaxies with spectra that have VESPA SFH
reconstructions by Tojeiro et al. (2009). Excluding 
 galaxies beyond redshift
$z>0.4$ (beyond the detection limit of SNe Ia, see below),
several tens of objects with blueshifts of about $10$~km~s$^{-1}$ (likely stars
that were misidentified as galaxies), and
galaxies with VESPA  fits with $\chi^2>10$, 
leaves about 67,000 galaxies. Finally, we consider only galaxies
in the right ascension (RA) range  $-51^\circ<\alpha_{\rm J2000}<+57^\circ$  which Dilday et
al. (2010; D10) defined for the purpose of their SN rate measurements, and for which they 
performed detection-efficiency simulations (see below). This leaves
66,400 galaxies. About 600 of these are actually likely pairs of duplicate
SDSS spectra, taken at different dates, of the same galaxy (same
coordinates to within $3''$ and same redshifts to within $300$~km~s$^{-1}$).
The resulting overcount 
of the number of monitored galaxies by about 300 (i.e. by a factor of 0.5\%), is of 
no consequence to our analysis. 
We consider these $\sim 66,000$ galaxies as the spectral galaxy
sample that was monitored for SNe in SDSS2.

As in B10 and M11, for the majority of the SDSS galaxies, it is practical to separate
the SFHs into no more than four time bins:
$0-70$~Myr, $70-420$~Myr, 
420~Myr$-$2.4~Gyr, and $>2.4$~Gyr, corresponding to the bins labeled 
24, 25, 26,
and 27 in Tojeiro et al. (2009). Furthermore (also as in B10 and M11), it is clear that 
VESPA often does not separate reliably the SFHs in the first two bins, so we combine
the first two time bins into a single bin, of $0-420$~Myr. We use
the VESPA SFH reconstructions based on 
the Maraston (2005) spectral synthesis models with a single dust component.

We scale down the 
stellar masses listed in the VESPA database by
a factor of 0.55, for the following reasons. 
To account for the limited $3''$ SDSS fibre aperture, Tojeiro et
al. (2009)
scaled up the formed masses reconstructed from the SDSS spectrum of
each galaxy, based on
the difference between the $z$-band magnitudes for the central $3''$ 
and the total Petrosian magnitudes (Blanton et al. 2001), where both of these 
magnitudes are based on the SDSS imaging.   
However, the spectral magnitudes, obtained by integrating the
SDSS spectrum over the relevant filter curves, and which matched 
the fibre magnitudes in SDSS Data Release 5 (DR5), no longer do so in 
the later data releases, including DR7 on whose spectra 
the reconstructed VESPA masses are based 
(R. Tojeiro 2010, private communication).  
This recalibration leads (Adelman-McCarthy et al. 2008) 
to a 0.35~mag shift (a factor of 0.72).
However, a further factor of 0.75 to 0.80 is required 
(R. Tojeiro 2010, private communication)
in order to make
the VESPA galaxy masses match the mass estimates of the same
galaxies based on pre-DR6 SDSS
spectroscopic calibrations (Gallazzi et al. 2005; Brinchman et al. 2003).
Apparently, this is the result of the DR7 spectral calibration
 being, on average, slightly redder than pre-DR6 spectra at
 wavelengths $\gtrsim 6500$~\AA. This, in turn leads to higher mass
estimates in DR7. Furthermore, we have found a similar factor 0.75 scaling
by comparing the VESPA masses available for 1282 galaxies in the sample of 
Mannucci et al. (2005) to the masses of those galaxies based
on their $B$ and $K$-band photometry, and using
the relations of Bell \& de Jong (2001). Finally, a similar factor
of 0.75 is suggested by comparing VESPA masses and PEGASE (Rocca-Volmerange et
al. 2011) masses, obtained for a sample of 13 galaxies by Neill et
al. (2009), by fitting their spectral energy distributions to broad-band
photometry. It is possible that, in fact, it is the DR7 spectra
and their VESPA reconstructions that give the more accurate stellar
masses, in which case it would be the previous SN rates per unit mass and
DTDs (e.g. Mannucci et al. 2005; Li et al. 2011; M11) that would have
to be scaled down by a factor 0.75. 
As this is still unclear, we have chosen here to    
apply 
this additional factor of 0.75, resulting in a full factor 0.55 
correction, to the Stripe 82 galaxy VESPA SFHs. The full factor of 0.55 
was also applied to the
galaxy SFHs in M11, which were also based on DR7 spectra. 
Neither part of this correction, however, was applied in
B10, and as a result the DTD 
values in B10 will be about a factor of 2 lower than here and in M11, 
just for this reason. 
 
The VESPA galaxies in Stripe 82 are our chosen sample of galaxies
monitored for SNe. 
For each galaxy, we therefore need to know the SN Ia detection 
efficiency and the effective visibility time. Dilday et al. (2008,
2010) have carried out detailed efficiency simulations for the purpose
of rate measurements, using fake SNe planted
in the SDSS2 images in the course of the survey. These images were then
searched for SN candidates, and subjected to detection and classification criteria, just
like the real SNe. We adopt the fiducial 
SN Ia detection efficiency, as a function of redshift, as given in
figure~8 of D10. 
We approximate this function with a piecewise linear
dependence --  a constant for $z<0.175$:
$\epsilon(z<0.175)=0.72$; and a linear decrease to $\epsilon=0$
between $z=0.175$ and $z=0.4$: $\epsilon(z)=-3.2 z +1.28$. This approximation matches the
empirical curve to a few percent at all redshifts. The $< 100$\%
efficiency, even at low redshifts where all SNe Ia are detected by the
survey, mainly reflects the selection criteria of the survey, which
required that SNe included in the final sample be observed near maximum light and also well after maximum
light. This effectively rejects SNe that exploded very near the beginning
and near the end of every observing season (see Dilday et al. 2008, and further below).
This detection
efficiency function differs from the one used in
B10. In B10, a particular functional form for $\epsilon(z)$ was assumed, with a number of
free parameters. These parameters were then tuned such that mock
samples of SNe Ia exploding in the sample galaxies, for an assumed
DTD, would have a
similar redshift distribution to that of the real SN Ia sample. 
Overall, the B10 efficiency function assumes  a lower assumed
detection efficiency, which leads to increased DTD values.   

SDSS2 was a ``rolling survey'',
in which the observer-frame visibility time for each monitored galaxy
was, 
in principle, just the  
the duration of the survey, 269 days (D10). The loss of
visibility time at the ``edges'' of each observing season is already 
accounted for in the detection efficiency function, as discussed
above. In the rest frame  of a galaxy at redshift $z$, the SN visibility time 
is reduced by a
factor $1+z$ due to cosmological time dilation. This reduction was
inadvertently omitted in B10 (although the tuning of the detection efficiency
parameters in order the match the SN redshift distribution may have
partly accounted for this effect).

\subsection{Supernova sample}
We turn next to the SN sample.
About one-thousand likely SN candidates were discovered in the course of SDSS2, 
and some of these candidates were then followed up with spectroscopic observations. 
 We select our SN Ia sample from among two subsamples
defined by D10 and one subsample defined by Sako et al. (2011;
S11). From these three subsamples, we 
associate between SNe and Stripe 82 galaxy hosts. D10 imposed selection criteria
for inclusion of SN Ia candidates in their samples, criteria that were applied also in 
their detection and
classification simulations and are therefore reflected in their
detection efficiency function (see above). They included:\\
1. At least one photometric measurement with signal-to-noise ratio
$S/N>5$ in each of three bands, $g$, $r$, and $i$;\\
2. A photometric observation at least 2 days (in the SN rest frame) before maximum of the
best-fit light-curve;\\
3.  A photometric observation at least 10 days (rest frame) after maximum of the
best-fit light-curve;\\
4. A good fit to a SN Ia light curve based on the MLCS2k2 program
(Jha et al. 2007), with fit criteria detailed in D10; and\\
5. RA in the range $-51^\circ<\alpha_{\rm J2000}<+57^\circ$.\\
With these criteria, D10 obtained a subsample of 312 spectroscopically
confirmed SNe Ia, listed in their table 2. A second subsample,
consisting of an
additional 148 likely SNe Ia satisfying the above criteria, was obtained
by D10 (and listed in their table 3) by collecting events that do not have spectroscopic confirmation  
as SNe Ia, but are very likely SNe Ia based on their photometric
light curve analysis (criterion 4, above), and which are associated with a 
host galaxy at a spectroscopic redshift consistent with the
photometric redshift from the light curve analysis.

S11 also assembled a ``photometric'' sample of 210 SDSS2 SNe~Ia 
(listed in their table 3), 
lacking spectroscopic confirmation of the SN 
but with host galaxies having spectroscopic
redshifts. The SN Ia classifications of S11 were 
based on photometric criteria that are not
identical, but quite similar to those of D10:\\
1. At least one photometric measurement with signal-to-noise ratio
$S/N>5$ in two of three bands, $g$, $r$, and $i$;\\
2. A photometric observation at $-5<t<+5$~days (rest frame) around
maximum of the
best-fit light-curve;\\
3.  A photometric observation at $+5<t<+15$~days (rest frame) after maximum of the
best-fit light-curve; and \\
4. A good fit to a SN Ia light curve based on the PSNID program
(Sako et al. 2008), with criteria detailed in S11.\\
Applying the D10 criterion on RA (No. 5, above) to the S11 sample
reduces it from 210 to 204 SNe.
There are 107 SNe common to these D10 and S11 photometric SN Ia samples.    

We have searched for associations between the Stripe 82 VESPA galaxies and 
the SNe Ia from these three
 subsamples (the D10 spectroscopically confirmed sample, and the D10 and
 S11 photometric samples),
 by requiring a
projected physical SN-host separation of $<30$~kpc, and a difference
of $\Delta z < 0.001$ between the spectroscopic galaxy redshift and
 the SN redshift (spectroscopic for the confirmed sample, photometric,
 based on the light-curve fitting, for the two other samples). Among the
 matches found in practice, except for seven cases, the redshift 
difference is actually $\Delta z < 0.0005$.
The SN-host separation is always $<26$~kpc, and except for 11 cases
it is $<20$~kpc. 
We find 61 galaxy matches to the spectrally confirmed SN Ia sample.
All but one of these events, the SN with SDSS designation 15467, have
IAU designations. We further match 32 SNe from the D10 photometric
sample, and 62 SNe from the S11 photometric sample, after imposing on 
the latter the D10 RA criteria (which removes 3 SNe). Of the photometric SNe, 23 are common
to the two samples, resulting in 71 unique photometric SNe.
Together with the spectroscopic SN Ia sample, we therefore have a
final sample of 132 SNe Ia that were hosted by the Stripe 82 VESPA
galaxies (in the restricted RA range) during the SDSS2 SN survey, and were detected 
and classified with the efficiency estimated by
D10. Contamination of the photometric samples by non-SN Ia events is estimated
at 3\% (D10) and 6\% (S11). We therefore expect that only three or so
of the 132 SNe are misclassified core-collapse SNe. Table 1 lists
the adopted SN Ia sample.

The SN Ia sample assembled by B10 in their DTD recovery analysis is
similar              in size (101 SNe) to the present sample (132 SNe),
but quite different in composition. B10
selected from the SN list available in the SDSS2 website the events that were
listed as
spectroscopically confirmed SNe Ia with IAU designations, and with
light curves of sufficient quality for the light-curve analysis done by B10.
However, this SN compilation is not complete. 
The D10 sample of spectrally 
confirmed SNe has 12 SNe with Stripe-82 host-galaxy matches that are
not included in the B10 sample. More importantly, the B10 sample omits the 
many SNe that were not observed spectroscopically, but are nonetheless
{\it bona-fide} 
SNe Ia, as confirmed by their photometric analysis. On the other hand,
B10 did include four SNe from the 2004 observing season, 
before the full SDSS2 SN survey.
In the end, only 50 SNe are common to B10 and to the present sample. 
Finally, as already noted,
this compilation of SNe did not have an associated detection
and classification efficiency function, which B10 therefore
estimated by reproducing the SN host redshift distribution.

\begin{table*}
%\scriptsize
\begin{minipage}{150mm}
\caption{Supernova and host galaxy sample}
\begin{tabular}{r|r|r|c|r|r|c|r|r|l}
\hline
\hline
\multicolumn{4}{c}{Supernova}&
\multicolumn{3}{c}{Host galaxy}&
\multicolumn{2}{c}{SN-host sep.}&
IAU\\
{ID} &
{RA} &
{Dec} &
{$z$} &
{RA} &
{Dec} &
{$z$} &
$''$ &
kpc&
ID \\
{(1)} &
{(2)} &
{(3)} &
{(4)} &
{(5)} &
{(6)} &
{(7)} &
{(8)} &
{(9)} &
{(10)}\\
\hline
\multicolumn{4}{l}{Spectroscopically confirmed SNe}& &&&&&\\
\hline
   762&   15.53518&   -0.87907&    0.1910&   15.53606&   -0.87966&    0.19154&  3.82& 12.18 &   2005eg\\
  1032&   46.79565&    1.11952&    0.1300&   46.79592&    1.12000&    0.12978&  1.97&  4.56 &   2005ez\\
  1112&  339.01761&   -0.37527&    0.2580&  339.01691&   -0.37489&    0.25777&  2.87& 11.47 &   2005fg\\
  1371&  349.37375&    0.42929&    0.1190&  349.37369&    0.42966&    0.11906&  1.36&  2.92 &   2005fh\\
  2561&   46.34335&    0.85829&    0.1180&   46.34430&    0.85972&    0.11841&  6.19& 13.23 &   2005fv\\
  2689&   24.90027&   -0.75879&    0.1620&   24.90004&   -0.75796&    0.16153&  3.10&  8.62 &   2005fa\\
  2992&   55.49692&   -0.78269&    0.1270&   55.49731&   -0.78294&    0.12656&  1.67&  3.79 &   2005gp\\
  3241&  312.65143&   -0.35416&    0.2590&  312.65280&   -0.35423&    0.25900&  4.95& 19.87 &   2005gh\\
  3592&   19.05240&    0.79183&    0.0870&   19.05289&    0.79061&    0.08661&  4.74&  7.70 &   2005gb\\
  3901&   14.85039&    0.00252&    0.0630&   14.85049&    0.00266&    0.06286&  0.61&  0.74 &   2005ho\\
  6057&   52.55355&   -0.97467&    0.0670&   52.55368&   -0.97448&    0.06706&  0.83&  1.06 &   2005if\\
  6295&   23.67295&   -0.60544&    0.0800&   23.67429&   -0.60420&    0.07961&  6.58&  9.89 &   2005js\\
  6406&   46.08860&   -1.06301&    0.1250&   46.08863&   -1.06312&    0.12463&  0.41&  0.91 &   2005ij\\
  6558&   21.70165&   -1.23814&    0.0570&   21.70190&   -1.23815&    0.05740&  0.90&  1.00 &   2005hj\\
  7876&   19.18237&    0.79452&    0.0760&   19.18279&    0.79361&    0.07637&  3.59&  5.20 &   2005ir\\
 12780&  322.15454&    1.22804&    0.0490&  322.15671&    1.23017&    0.04944& 10.94& 10.58 &   2006eq\\
 12856&  332.86542&    0.75589&    0.1720&  332.86539&    0.75559&    0.17171&  1.09&  3.17 &   2006fl\\
 12874&  353.96448&   -0.17724&    0.2450&  353.96329&   -0.17659&    0.24502&  4.89& 18.83 &   2006fb\\
 12950&  351.66742&   -0.84032&    0.0830&  351.66730&   -0.84061&    0.08271&  1.13&  1.76 &   2006fy\\
 13070&  357.78500&   -0.74645&    0.1990&  357.78491&   -0.74657&    0.19855&  0.56&  1.83 &   2006fu\\
 13072&  334.95929&    0.02439&    0.2310&  334.96069&    0.02368&    0.23062&  5.67& 20.86 &   2006fi\\
 13354&   27.56474&   -0.88735&    0.1580&   27.56472&   -0.88671&    0.15759&  2.30&  6.26 &   2006hr\\
 13511&   40.61224&   -0.79419&    0.2380&   40.61127&   -0.79422&    0.23758&  3.49& 13.14 &   2006hh\\
 14279&   18.48869&    0.37156&    0.0450&   18.48826&    0.37143&    0.04543&  1.62&  1.45 &   2006hx\\
 14284&   49.04921&   -0.60105&    0.1810&   49.04937&   -0.60099&    0.18109&  0.61&  1.87 &   2006ib\\
 14377&   48.26427&   -0.47171&    0.1390&   48.26377&   -0.47174&    0.13939&  1.80&  4.43 &   2006hw\\
 15129&  318.90228&   -0.32151&    0.1990&  318.90210&   -0.32171&    0.19846&  0.97&  3.19 &   2006kq\\
 15136&  351.16245&   -0.71842&    0.1490&  351.16220&   -0.71793&    0.14867&  1.98&  5.14 &   2006ju\\
 15161&   35.84275&    0.81893&    0.2500&   35.84263&    0.81904&    0.24954&  0.59&  2.30 &   2006jw\\
 15222&    2.85320&    0.70267&    0.1990&    2.85241&    0.70200&    0.19933&  3.72& 12.26 &   2006jz\\
 15234&   16.95826&    0.82809&    0.1360&   16.95807&    0.82859&    0.13630&  1.93&  4.66 &   2006kd\\
 15421&   33.74157&    0.60240&    0.1850&   33.74128&    0.60272&    0.18537&  1.57&  4.86 &   2006kw\\
 15425&   55.56103&    0.47820&    0.1600&   55.56109&    0.47835&    0.16003&  0.59&  1.64 &   2006kx\\
 15443&   49.86736&   -0.31813&    0.1820&   49.86745&   -0.31799&    0.18203&  0.61&  1.86 &   2006lb\\
 15467&  320.01984&   -0.17753&    0.2100&  320.02011&   -0.17735&    0.21042&  1.18&  4.04 &  \nodata\\
 15648&  313.71829&   -0.19488&    0.1750&  313.71881&   -0.19581&    0.17499&  3.84& 11.41 &   2006ni\\
 16211&  348.16397&    0.26679&    0.3110&  348.16290&    0.26598&    0.31091&  4.83& 22.06 &   2006nm\\
 17186&   31.61272&   -0.89963&    0.0800&   31.61644&   -0.89802&    0.07980& 14.59& 21.98 &   2007hx\\
 17332&   43.77335&   -0.14750&    0.1830&   43.77249&   -0.14769&    0.18299&  3.16&  9.73 &   2007jk\\
 17340&   41.21201&    0.36474&    0.2570&   41.21338&    0.36531&    0.25692&  5.35& 21.32 &   2007kl\\
 17366&  315.78723&   -1.02926&    0.1390&  315.78500&   -1.03118&    0.13934& 10.58& 25.99 &   2007hz\\
 17497&   37.13649&   -1.04221&    0.1450&   37.13650&   -1.04286&    0.14478&  2.34&  5.94 &   2007jt\\
 17629&   30.63637&   -1.08924&    0.1370&   30.63648&   -1.08995&    0.13691&  2.60&  6.30 &   2007jw\\
 17784&   52.46166&    0.05676&    0.0370&   52.46180&    0.05444&    0.03714&  8.37&  6.17 &   2007jg\\
 17880&   44.97227&    1.16063&    0.0730&   44.97361&    1.16003&    0.07275&  5.29&  7.32 &   2007jd\\
 18030&    4.93285&   -0.40022&    0.1560&    4.93321&   -0.40009&    0.15647&  1.37&  3.71 &   2007kq\\
 18298&   18.26664&   -0.54014&    0.1200&   18.26680&   -0.54002&    0.11980&  0.71&  1.54 &   2007li\\
 18612&   12.28788&    0.59688&    0.1150&   12.28801&    0.59660&    0.11505&  1.10&  2.29 &   2007lc\\
 18835&   53.68503&    0.35546&    0.1230&   53.68539&    0.35550&    0.12324&  1.30&  2.88 &   2007mj\\
 18855&   48.63231&    0.26975&    0.1280&   48.63386&    0.26887&    0.12783&  6.42& 14.67 &   2007mh\\
 18890&   16.44440&   -0.75890&    0.0660&   16.44335&   -0.75947&    0.06645&  4.31& 22.61 &   2007mm\\
 18903&   12.25138&   -0.32410&    0.1560&   12.25120&   -0.32326&    0.15635&  3.11&  8.40 &   2007lr\\
 19155&   31.26644&    0.17447&    0.0770&   31.26481&    0.17512&    0.07704&  6.31&  9.21 &   2007mn\\
 19353&   43.11432&    0.25177&    0.1540&   43.11329&    0.25174&    0.15397&  3.71&  9.91 &   2007nj\\
 19616&   37.10098&    0.18458&    0.1660&   37.09966&    0.18600&    0.16538&  6.99& 19.80 &   2007ok\\
 19794&  359.31897&    0.24923&    0.2970&  359.31909&    0.24849&    0.29708&  2.69& 11.90 &   2007oz\\
 19968&   24.34869&   -0.31209&    0.0560&   24.34907&   -0.31173&    0.05604&  1.89&  2.06 &   2007ol\\
 19969&   31.91040&   -0.32408&    0.1750&   31.90982&   -0.32403&    0.17527&  2.09&  6.22 &   2007pt\\
 20064&  358.58612&   -0.91773&    0.1050&  358.58630&   -0.91722&    0.10499&  1.96&  3.78 &   2007om\\
 20084&  347.97513&   -0.57817&    0.0910&  347.97641&   -0.57909&    0.09123&  5.67& 17.08 &   2007pd\\
 20350&  312.80576&   -0.95590&    0.1300&  312.80679&   -0.95778&    0.12946&  7.74& 17.87 &   2007ph\\
\end{tabular}
\label{table1}
\end{minipage}
\end{table*}

%%%%%%%%%%%%%%%%

\begin{table*}
%\scriptsize
\begin{minipage}{150mm}
%\caption{(Continued)}
\begin{tabular}{r|r|r|c|r|r|c|r|r|l}
\hline
\hline
\multicolumn{4}{c}{Supernova}&
\multicolumn{3}{c}{Host galaxy}&
\multicolumn{2}{c}{SN-host sep.}&
IAU\\
{ID} &
{RA} &
{Dec} &
{$z$} &
{RA} &
{Dec} &
{$z$} &
$''$ &
kpc&
ID \\
{(1)} &
{(2)} &
{(3)} &
{(4)} &
{(5)} &
{(6)} &
{(7)} &
{(8)} &
{(9)} &
{(10)} \\
\hline
\multicolumn{4}{l}{Photometrically classified SNe}& &&&&&\\
\hline
  1415&    6.10647&    0.59921&    0.2120&    6.10649&    0.59902&    0.21195&  0.69&  2.40&  	      \\
  1748&  353.11215&   -0.48250&    0.3397&  353.11179&   -0.48223&    0.33975&  1.63&  7.91&  	      \\
  4019&    1.26181&    1.14542&    0.1814&    1.26194&    1.14637&    0.18140&  3.45& 10.53&  	      \\
  4651&   37.37555&   -0.74726&    0.1520&   37.37570&   -0.74753&    0.15179&  1.10&  2.91&  	      \\
  4690&   32.92946&    0.68817&    0.2000&   32.93032&    0.68770&    0.19918&  3.53& 11.63&  	      \\
  6213&  344.10583&   -0.44993&    0.1094&  344.10580&   -0.44998&    0.10930&  0.19&  0.38&  	      \\
  6332&  325.96484&   -0.71813&    0.1522&  325.96552&   -0.71828&    0.15223&  2.48&  6.56&  	      \\
  6638&   45.03818&   -1.23639&    0.3257&   45.03856&   -1.23757&    0.32580&  4.46& 21.00&  	      \\
  6851&   52.10445&   -0.04860&    0.3050&   52.10456&   -0.04880&    0.30503&  0.82&  3.68&  	      \\
  7350&    7.56355&   -0.78711&    0.1555&    7.56280&   -0.78721&    0.15545&  2.74&  7.38&  	      \\
  7431&  340.95438&   -0.27500&    0.3500&  340.95459&   -0.27462&    0.35019&  1.56&  7.71&  	      \\
  7479&    7.22581&   -0.40948&    0.2272&    7.22587&   -0.40948&    0.22726&  0.24&  0.88&  	      \\
  8004&  347.52853&   -0.55806&    0.3514&  347.52869&   -0.55806&    0.35138&  0.55&  2.72&  	      \\
  8195&  331.00635&   -0.89569&    0.2690&  331.00641&   -0.89567&    0.26886&  0.23&  0.93&  	      \\
  8280&    8.57363&    0.79595&    0.1850&    8.57315&    0.79553&    0.18506&  2.30&  7.14&  	      \\
  8555&    2.91543&   -0.41497&    0.1980&    2.91559&   -0.41505&    0.19771&  0.65&  2.12&  	      \\
  8888&    5.16572&    0.32054&    0.3985&    5.16573&    0.32142&    0.39865&  3.16& 16.92&  	      \\
  9117&   46.90165&    0.98827&    0.2720&   46.90008&    0.98841&    0.27215&  5.67& 23.57&  	      \\
  9133&   16.64246&    0.46087&    0.2672&   16.64254&    0.46055&    0.26722&  1.17&  4.80&  	      \\
  9558&   15.98495&    0.38134&    0.3908&   15.98441&    0.38095&    0.39069&  2.39& 12.66&  	      \\
  9633&   33.91829&    1.09527&    0.1958&   33.91864&    1.09519&    0.19575&  1.31&  4.24&  	      \\
  9739&  323.69379&   -0.87893&    0.1200&  323.69470&   -0.87905&    0.12022&  3.33&  7.21&  	      \\
 10690&  347.49985&    1.08225&    0.3123&  347.49991&    1.08216&    0.31223&  0.38&  1.74&  	      \\
 11306&   56.73846&   -0.51832&    0.2740&   56.73822&   -0.51760&    0.27374&  2.72& 11.36&  	      \\
 11311&   47.01525&    0.43353&    0.2046&   47.01541&    0.43347&    0.20455&  0.61&  2.04&  	      \\
 12310&  351.52161&    1.02211&    0.1510&  351.52328&    1.02083&    0.15097&  7.61& 20.00&  	      \\
 13071&  358.61783&   -0.71871&    0.1789&  358.61801&   -0.71874&    0.17886&  0.67&  2.02&  	      \\
 13458&   16.46333&   -0.25047&    0.3191&   16.46215&   -0.24955&    0.31911&  5.37& 24.95&  	      \\
 13545&   52.34275&    0.59634&    0.2141&   52.34276&    0.59630&    0.21411&  0.16&  0.55&  	      \\
 13633&    4.66560&    0.00587&    0.3880&    4.66536&    0.00548&    0.38773&  1.66&  8.75&  	      \\
 14340&  345.82657&   -0.85538&    0.2770&  345.82669&   -0.85525&    0.27744&  0.65&  2.75&  	      \\
 14398&   50.71037&    0.27259&    0.1183&   50.71052&    0.27282&    0.11831&  0.97&  2.08&  	      \\
 14961&   15.91944&    0.93097&    0.3700&   15.91944&    0.93131&    0.37051&  1.21&  6.19&  	      \\
 15262&  342.43274&   -0.40416&    0.2474&  342.43271&   -0.40420&    0.24732&  0.19&  0.75&  	      \\
 15303&  350.50882&    0.54095&    0.2340&  350.50891&    0.54092&    0.23429&  0.34&  1.28&  	      \\
 15454&  327.85934&   -0.84815&    0.3830&  327.85870&   -0.84918&    0.38280&  4.36& 22.82&  	      \\
 15580&   54.49311&    0.48326&    0.3225&   54.49328&    0.48326&    0.32257&  0.60&  2.83&  	      \\
 15727&  351.40002&   -0.01417&    0.1055&  351.40039&   -0.01453&    0.10548&  1.86&  3.59&  	      \\
 15765&   32.84710&    0.24601&    0.3050&   32.84801&    0.24622&    0.30485&  3.37& 15.17&  	      \\
 15950&    6.18888&    0.98369&    0.2204&    6.18881&    0.98408&    0.22049&  1.39&  4.94&  	      \\
 15971&   40.11259&    0.52619&    0.3160&   40.11262&    0.52628&    0.31591&  0.33&  1.52&  	      \\
 16136&   33.48586&   -0.73159&    0.3427&   33.48594&   -0.73169&    0.34267&  0.47&  2.29&  	      \\
 16163&   31.49926&   -0.85571&    0.1549&   31.49916&   -0.85584&    0.15525&  0.58&  1.57&  	      \\
 16462&   17.04058&   -0.38642&    0.2446&   17.04063&   -0.38594&    0.24460&  1.74&  6.70&  	      \\
 16563&  343.30557&   -0.09531&    0.3453&  343.30591&   -0.09544&    0.34527&  1.29&  6.34&  	      \\
 16768&  322.70056&    0.69283&    0.1690&  322.70181&    0.69260&    0.16895&  4.58& 13.20&  	      \\
 17206&   45.98524&    0.72816&    0.1560&   45.98590&    0.72821&    0.15640&  2.38&  6.45&  	      \\
 17434&   18.43727&   -0.07290&    0.1790&   18.43766&   -0.07295&    0.17873&  1.42&  4.29&  	      \\
 17958&   34.43497&   -0.71299&    0.2760&   34.43525&   -0.71297&    0.27600&  1.00&  4.22&  	      \\
 18047&   22.07419&   -0.65805&    0.3590&   22.07432&   -0.65857&    0.35915&  1.93&  9.68&  	      \\
 18201&   47.31259&   -0.64449&    0.2931&   47.31229&   -0.64498&    0.29308&  2.09&  9.17&  	      \\
 18224&  348.17462&   -0.31281&    0.3392&  348.17480&   -0.31286&    0.33913&  0.68&  3.28&  	      \\
 18273&  334.73901&   -0.63039&    0.3163&  334.73950&   -0.63028&    0.31634&  1.81&  8.35&  	      \\
 18630&  347.97998&   -0.26443&    0.3590&  347.98029&   -0.26411&    0.35948&  1.59&  8.01&  	      \\
 18647&  322.89191&   -0.30387&    0.2130&  322.89319&   -0.30288&    0.21275&  5.84& 20.20&  	      \\
 19048&  326.12454&    0.59555&    0.1368&  326.12411&    0.59578&    0.13677&  1.75&  4.24&  	      \\
 19090&  357.16226&   -0.40642&    0.3120&  357.16229&   -0.40647&    0.31202&  0.22&  1.02&  	      \\
 19317&  310.45483&    1.06483&    0.1787&  310.45459&    1.06486&    0.17873&  0.88&  2.67&  	      \\
 20047&  326.59094&    0.63138&    0.3739&  326.59091&    0.63109&    0.37388&  1.03&  5.33&  	      \\
 20141&  357.54199&   -0.52391&    0.3420&  357.54120&   -0.52447&    0.34144&  3.49& 16.98&  	      \\
 20314&    4.17917&    0.72267&    0.2100&    4.17927&    0.72254&    0.20998&  0.56&  1.92&  	      \\
\end{tabular}
\label{table1cont}
\end{minipage}
\end{table*}

%%%%%%%%%%%%%%%%
\begin{table*}
%\scriptsize
\begin{minipage}{150mm}
%\caption{(Continued)}
\begin{tabular}{r|r|r|c|r|r|c|r|r|l}
\hline
\hline
\multicolumn{4}{c}{Supernova}&
\multicolumn{3}{c}{Host galaxy}&
\multicolumn{2}{c}{SN-host sep.}&
IAU\\
{ID} &
{RA} &
{Dec} &
{$z$} &
{RA} &
{Dec} &
{$z$} &
$''$ &
kpc&
ID \\
{(1)} &
{(2)} &
{(3)} &
{(4)} &
{(5)} &
{(6)} &
{(7)} &
{(8)} &
{(9)} &
{(10)}\\
\hline
 20331&    7.53944 &   1.24604  &  0.1845&    7.53964&    1.24591  &  0.18449 & 0.84 & 2.59&  	      \\
 20480&  357.27420 &   0.91829  &  0.1678&  357.27429&    0.91820  &  0.16776 & 0.47 & 1.35&  	      \\
 20626&    8.47575 &  -0.59308  &  0.2760&    8.47590&   -0.59297  &  0.27628 & 0.67 & 2.80&  	      \\
 20721&  323.18481 &  -0.62219  &  0.2120&  323.18481&   -0.62275  &  0.21183 & 2.01 & 6.94&  	      \\
 20726&   42.27992 &  -0.15946  &  0.3201&   42.27989&   -0.15947  &  0.32015 & 0.13 & 0.62&  	      \\
 20787&   50.51086 &  -0.44245  &  0.2707&   50.51080&   -0.44254  &  0.27066 & 0.40 & 1.67&  	      \\
 20788&   51.67339 &  -0.47783  &  0.3940&   51.67345&   -0.47777  &  0.39347 & 0.31 & 1.67&  	      \\
 21709&  326.52615 &  -1.04962  &  0.1590&  326.52771&   -1.04932  &  0.15900 & 5.71 &15.66&  	      \\
 21872&    7.61328 &  -0.59137  &  0.2286&    7.61338&   -0.59148  &  0.22854 & 0.52 & 1.91&  	      \\
 22006&   48.78285 &  -0.96254  &  0.3956&   48.78292&   -0.96260  &  0.39556 & 0.32 & 1.72&          \\
\hline										      
\end{tabular}

{Column header explanations: \\
(1)- SDSS-II supernova identifier; 
\\
(2-4) - Supernova right ascension and declination, J2000, in decimal degrees, and redshift;\\ 
(5-7) - Associated host galaxy right ascension and declination, J2000, in decimal degrees, and redshift;\\ 
(8-9) - Supernova-host separation, in arcseconds, and in projected kpc (assuming
cosmological parameters $h=0.7$, $\Omega_m=0.3$,
$\Omega_\Lambda=0.7$);\\ 
(10) - Supernova IAU designation (for spectroscopically confirmed SNe only).
}
\label{table1cont}
\end{minipage}
\end{table*}

\section{DTD recovery method}
We briefly repeat here the description of the DTD recovery method of M11.
For a sample of $N$ galaxies,
the SN rate in the $i$th galaxy observed at cosmic 
time $t$ is given by the convolution
\begin{equation}
\label{convolution}
r_i(t)=\int_0^{t} S_i(t-\tau)\Psi (\tau) d\tau ,
\end{equation}
where: $S_i(t)$ is the star-formation rate versus cosmic time, of the
$i$th galaxy (stellar mass formed per unit
time); $\Psi (\tau)$ is the DTD (SNe per unit time per unit stellar mass
formed); and the integration is from the Big Bang ($t=0$) to the time
of observation. We assume that the
DTD is a universal function: it is the same in all galaxies,
independent of environment,
metallicity, and cosmic time --- 
neglecting such possible dependences (e.g.,
variations in the IMF with
cosmic times or with environment would also lead to a variable DTD).
The DTD is recovered by inverting
a linear, discretized version of Eq.~\ref{convolution}.
The SFHs of the
$i=1,2,...,N$ galaxies 
monitored as part of 
a SN survey are known  with a temporal resolution that permits binning the 
stellar mass formed in each galaxy 
into $j=1,2,..,K$ discrete (but not necessarily equal) time bins, where 
increasing $j$ corresponds to increasing lookback time. 
For the $i$th galaxy
in the survey, the stellar mass formed in the $j$th time bin is
$m_{ij}$. 
The mean of the DTD over the $j$th bin (corresponding to a
delay range equal to the lookback-time range of the $j$th bin in the
SFH) is $\Psi_j$. Then  the integration in Eq.~\ref{convolution} can
be approximated as a sum,
\begin{equation}
\label{discreteconvolution}
r_i\approx\sum_{j=1}^K m_{ij}\Psi_j  ,
\end{equation}
where $r_i$, the SN rate in a given galaxy, is measured at a
particular cosmic time, corresponding to the galaxy redshift, $z_i$.
 Given a survey of $N$ galaxies, each with 
an observed SN rate, $r_i$, and a known binned SFH,
$m_{ij}$, one could, in principle, algebraically 
invert this set of linear equations and 
recover the best-fit parameters describing the binned DTD: 
${\bf \Psi}=(\Psi_1,
\Psi_2,...,\Psi_K)$.

In practice, on human timescales SNe in a given
galaxy are rare events ($r_i\ll 1 ~{\rm yr}^{-1}$). Supernova surveys
therefore monitor many galaxies. For a given model  DTD,
${\bf \Psi}$, the $i$th galaxy
will have an expected number of SNe
\begin{equation}
\label{riti}
\lambda_i=r_i t_i \epsilon_i,
\end{equation}
where $t_i\epsilon_i$ is the effective visibility time 
during which a SN of a particular type 
would have been visible. In the present context of the SDSS2 rolling survey,
$t_i=269~{\rm d}/(1+z_i)$, and $\epsilon_i$ is the detection efficiency
based on the D10 simulations (which take into account
the actual on-target 
monitoring time, the SN detection limits at 
the redshift of the galaxy, the light-curve selection criteria, 
and the detection and classification efficiencies). Since $\lambda_i \ll 1$,
the number of SNe observed in the $i$th galaxy, $n_i$, obeys a Poisson
probability distribution with expectation value $\lambda_i$, 
\begin{equation}
P(n_i|\lambda_i)=(e^{-\lambda_i}\lambda_i^{n_i})/n_i ! ,
\end{equation}
where $n_i$ is 0 for most of
the galaxies, and 1 for some of the galaxies.

To deal with this Poisson aspect of the problem, M11 introduced
a non-parametric, maximum-likelihood method
to recover the DTD and its uncertainties, based on Cash (1979)
statistics. 
Considering a set of model DTDs, 
 the likelihood of a particular DTD, given the set of measurements
 $n_1,...,n_N$, is
\begin{equation}
L=\prod_{i=1}^N P(n_i|\lambda_i).
\end{equation}
More conveniently, the log of the likelihood is
\begin{equation}
\ln L =\sum_{i=1}^N \ln
P(n_i|\lambda_i)
%=-\sum_{i=1}^N{\lambda_i}
%+\sum_{i=1}^N\ln (\lambda_i^{n_i}/n_i !), 
%
%$$
=-\sum_{i=1}^N{\lambda_i}
+\sum_{n_i=1}\ln \lambda_i 
%+\sum_{n_i=2}\ln (\lambda_i^2/2) 
%+\sum_{n_i=3}\ln (\lambda_i^3/3!)+... ,
%$$
\end{equation}
%where the first sum on the right-hand side is over all galaxies,
%and the following sums include only the galaxies with one detected SN,
%with two detected SNe, etc. 
where only galaxies hosting SNe contribute to the second term.
The best-fitting model can be found by scanning the parameter space
of the vector ${\bf \Psi}$ for the value that maximizes the log-likelihood.
This procedure naturally allows restricting the DTD to have
only positive values, as physically required 
(a negative SN rate is meaningless).

The covariance matrix $C_{jk}$ of the uncertainties in the best-fit parameters 
can be found (e.g., Press et al. 1992) by
calculating the curvature matrix,
$$
\alpha_{jk}=\frac{1}{2}\frac{\partial^2\ln L}{\partial\Psi_j \partial\Psi_k}
=\sum_{i=1}^N\frac{\partial[\ln P(n_i|\lambda_i)]}{\partial\Psi_j}
\frac{\partial[\ln P(n_i|\lambda_i)]}{\partial\Psi_k}
$$
\begin{equation}
=\sum_{i=1}^N t_i^2(n_i/\lambda_i-1)^2 m_{ij} m_{ik} ,
\label{curvature}
\end{equation}
and inverting it,
\begin{equation}
[C]=[\alpha]^{-1} .
\end{equation}

%Because the values of the DTD are constrained to be positive, if the
%maximum-likelihood value of a DTD component, $\Psi_j$, is close to zero, 
%the square root of its variance, ${\sqrt{C_{jj}}}$, will not represent
%well its  $1\sigma$ uncertainty range. An alternative, 
%more reliable, procedure
%is to perform a Monte-Carlo simulation in which many mock surveys 
%are produced, each having the same galaxies, 
%SFHs, and visibility times as the real survey,
%and having expectation values $\lambda_i$ based on the best-fit DTD, but
%with the number of SNe in every galaxy, $n_i$, drawn from a Poisson
%distribution according to $\lambda_i$. The maximum-likelihood DTD,
%${\bf\Psi}$, is found for every realization. From the distribution 
% of the values of every component, $\Psi_j$, over 
%all the realizations, one can estimate the range encompassing, say,
%$\pm 34\%$ of the cases.

We have applied this DTD recovery method to the sample of 
monitored galaxies, and the SNe Ia that they hosted in SDSS2, as presented
in Section~\ref{datasection}.

\section{Results}
\subsection{Delay-time distribution}
\begin{figure*}
\begin{center}
\includegraphics[angle=-90, width=0.7\textwidth]{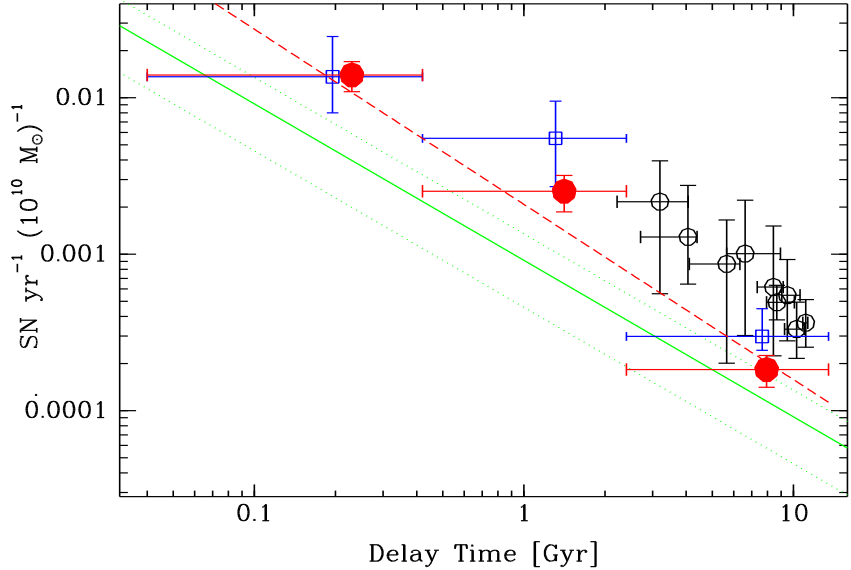}
\caption{Recovered SN~Ia DTD
for the SDSS2 SN sample. Filled red circles mark the
best-fit DTD values for each time bin, whose time range is indicated by the horizontal
 error bars. Vertical error bars show the  
Gaussian $1\sigma$ uncertainties. Red dashed line is the best
 power-law fit, with index $-1.12$, to the recovered DTD. 
Previous DTD measurements also shown are: M11
 analysis of the nearby LOSS sample (empty blue squares, with central
 values slightly shifted to the left, for clarity); Maoz et
 al. (2010) analysis based on SN Ia rates in galaxy clusters (empty
 black circles); and $t^{-1}$ power law DTD found by Graur et
 al. (2011), based on comparison of 
volumetric SN Ia rates and cosmic SFH (solid
 green line, with dotted lines marking the $1\sigma$ range). 
}
\label{sdss2dtdia}
\end{center}
\end{figure*}

Figure ~\ref{sdss2dtdia} shows the recovered DTD for our sample of 132 SNe Ia
discovered during the SDSS2 among the 66,000 Stripe-82 galaxies that have VESPA SFHs. The numerical 
values and uncertainties of the DTD are listed in Table~2.
Several aspects of the DTD are notable. First, we detect, at $>4\sigma$ significance,
a DTD signal in all three VESPA time bins, including the intermediate time bin, 
$0.42<t<2.4$~Gyr. B10 and M11 found highly significant detections in the prompt bin 
and in the delayed 
bin, but only $\sim 2\sigma$ DTD detections at intermediate delays (albeit consistent with the level
expected from a $t^{-1}$ DTD). Our new result provides the strongest support to date
for  previous indications for a continuous DTD
(Mannucci et al. 2006; Totani et al. 2008; Maoz et al. 2010). These previous DTD reconstructions,
however, in place of the actual SFHs, used luminosity-weighted galaxy ages (Totani et al. 2008) or
simple assumed SFHs (Mannucci et al. 2006, Maoz et al. 2010). 
The present result is the most significant and
precise detection of a DTD signal in all three of the time-delay bins defined here, 
in a DTD reconstruction that uses individual galaxy SFHs.

Also shown   
in Fig.~\ref{sdss2dtdia}, for comparison, are a number of other recent
DTD estimates. The value we find for the prompt DTD component,
$\Psi_1$, is in excellent agreement with the one found in M11. The
values of the intermediate and delayed components, $\Psi_2$ and
$\Psi_3$, respectively, are a factor $\sim 2$ lower than found by M11. 
In the case of $\Psi_2$, the difference is not statistically
significant, and furthermore the M11 value may have been overestimated due to
some residual ``leak'' of signal from $\Psi_1$ (see M11, and below). 
The difference in $\Psi_3$ values 
between M11 and the present work, however, is more significant, at the
$\sim 2\sigma$ level. Furthermore, both the current and the M11
 $\Psi_3$ measurements are lower than the DTD
values implied by a number of independent measurements of SN Ia rates
in galaxy clusters, as compiled in Maoz et al. (2010), and also shown 
in Fig.~\ref{sdss2dtdia}. On the other hand, as also shown in the
figure, our new $\Psi_3$
measurement is consistent in level with the DTD level at long delays
found by Graur et al. (2011) from comparison of volumetric SN Ia rates
to cosmic SFH. Our result thus revives previous speculation that the 
SN~Ia rate in old populations may be enhanced by a factor of a few
in cluster environments, compared to field environments (Sharon et
al. 2007; Mannucci et al. 2008). The intermediate $\Psi_3$ value found
in M11 could be due to a relatively large fraction of LOSS galaxies in 
clusters. Among all the LOSS galaxies, 20\% are in galaxy
clusters that are listed in the NED database, based on the same criteria as
applied in Mannucci et al. (2008; projected separation $<1.5$~Mpc, 
redshift difference $<1000$~km~s$^{-1}$). The fraction of the early-type LOSS
galaxies 
in clusters will be higher. 
Assuming the  SDSS2
galaxy sample, in contrast, is dominated by a field population, a
factor-few enhancement of the  SN Ia rate in clusters could lead
to the higher $\Psi_3$ value measured in M11, compared to here.
We defer to future work a more detailed examination of this
possibility regarding the LOSS sample.

It is important to note that the errors shown in Fig.~\ref{sdss2dtdia}
are statistical only. A systematic source of error, one that is
difficult to estimate, is associated with the SFH reconstruction of
each galaxy (see Tojeiro et al. 2009, M11, and Section 4.3, below).

\begin{table*}
%\scriptsize
\begin{minipage}{150mm}
\caption{Recovered Delay Time Distribution}
\begin{tabular}{l|c|c|c|c|c|c}
\hline
\hline
{Sample} &
{$N_{\rm gal}$} &
{$N_{\rm SN}$} &
{$\Psi_{1}$} &
{$\Psi_{2}$} &
{$\Psi_{3}$} &
{$N_{\rm SN}/M$} \\
{} &
{} &
{} &
{0--0.42~Gyr} &
{0.42--2.4~Gyr} &
{2.4--14~Gyr} &
{} \\
{(1)} &
{(2)} &
{(3)} &
{(4)} &
{(5)} &
{(6)} &
{(7)} \\
\hline
\hline

{\bf Full}&66,000&132&$140\pm30 $&$25.1\pm 6.3$&$1.83\pm0.42$&$0.00130\pm0.00015$\\
High $s$&66,000&67&$87\pm24 $&$10.5\pm 4.2$&$0.78\pm0.32$&$0.00066\pm0.00011$\\
Low $s$&66,000&45&$24\pm15 $&$5.7\pm 3.6$&$0.96\pm 0.27$&$0.00032\pm0.00007$\\
\hline
\end{tabular}

{Column header explanations: \\
(1)- Sample used to derive DTD: Full sample, high-stretch sample, or
  low-stretch sample.\\
(2) - Number of galaxies in sample.\\
(3) - Number of SNe in sample.\\
(4-6) - DTD rates and 68\% uncertainty ranges, 
in units of $10^{-14}$~SNe~yr$^{-1}\,{\rm M}_\odot^{-1}$.\\ 
%Upper limits, where
%given, correspond to a best fit of 0, and to the 95\% confidence
%limit.\\
(7) - Time-integrated DTD, in units of SNe~${\rm M}_\odot^{-1}$.
}
\label{table2}
\end{minipage}
\end{table*}

%power-law fit
We have used $\chi^2$ minimization to fit model power-law functions,
$\Psi(t)\propto t^{\beta}$, to the recovered binned DTD, with
the amplitude and the power-law index as free parameters. The model is
averaged over each time bin, to obtain the value that is compared to
the observed DTD value for that bin.
 The best-fitting power-law has
 an index of $\beta = -1.12\pm 0.08$. Our result augments a growing number of
 indications for a DTD that is a power law with a slope of $\sim -1$. 
However, Fig.~\ref{sdss2dtdia} suggests that a better description may 
be a function that steepens somewhat at long delays, compared to a
single power law. Further studying the DTD at such a level of detail
will require better temporal resolution. The present three-bin time
resolution is dictated by the spectral resolution and the
signal-to-noise ratio of the SDSS spectra.

\subsection{SN Ia production efficiency}
As discussed in Maoz \& Mannuci (2012), and references therein,
another interesting observable is the DTD normalization.
It can be discussed in the context of  
the integral of the DTD over the age of the universe, $t_0$,
\begin{equation}
N_{\rm SN}/M=\int_0^{t_0}\Psi(t)~ dt ,
\label{intddt}
\end{equation}
which gives the total number of
SNe that eventually explode, per unit stellar mass formed in a short
burst of star formation.
The integral in Eq.~\ref{intddt} can be approximated by 
a sum over the binned DTD,
\begin{equation}
N_{\rm SN}/M\approx\sum_{j=1,K}\Psi_j \Delta t_j.
\label{nmsum}
\end{equation}
 From the simulations described in M11, for declining power-law DTDs
represented with three time bins, Eq.~\ref{nmsum} typically underestimates
$N_{\rm SN}/M$ systematically by $\sim 10-20\%$. 

For the SDSS2 sample, we obtain 
$N_{\rm Ia}/M=0.00130\pm 0.00015~{\rm M}_\odot^{-1}$.
This is lower than the 
$N_{\rm Ia}/M=0.0020\pm 0.0006 ~{\rm M}_\odot^{-1}$ found in M11 for the LOSS sample,
although consistent with it, given the larger uncertainty 
of the M11 result. The present result is also consistent with a lower
estimate of $N_{\rm Ia}/M=0.0010\pm 0.0005 ~{\rm M}_\odot^{-1}$ found by Graur et al. (2011), from
comparison of volumetric SN Ia rate evolution and cosmic SFH. A very
recent analysis of the volumetric rates by Perrett et al. (2012) gives
$N_{\rm Ia}/M\sim 0.0006 - 0.0010 ~{\rm M}_\odot^{-1}$ (after
converting their result to a diet-Salpeter IMF), similar to the
Graur et al. (2011) result.
As was the case
when considering the DTD amplitude $\Psi_3$, above, the
estimates of $N_{\rm Ia}/M$ from galaxy clusters are higher 
when based on cluster SN Ia rates vs. redshift, and even more so when 
based on cluster iron mass content (Maoz et al. 2010; see Maoz \& Mannucci 2012, table 1).
For example, the $2\sigma$ lower limit  is
$N_{\rm Ia}/M>0.0017~{\rm M}_\odot^{-1}$ from SN Ia rates in clusters, and 
$N_{\rm Ia}/M>0.0034~{\rm M}_\odot^{-1}$ from the iron mass content. 
It remains to be seen whether this is the result of some  systematic error, or 
a real enhancement of SN Ia production in cluster environments 
(Sharon et al. 2007; Mannucci et al. 2008; Sand et al. 2011). 

The SN Ia number per formed stellar mass,  $N_{\rm Ia}/M$, can 
easily be related 
to the fraction, $\eta$, of stars in some initial mass range $[m_1,m_2]$,
that eventually explode as SNe~Ia:
\begin{equation}
\label{frac1}
\eta=\frac{N_{\rm Ia}}{M}\frac{\int_{0.1}^{100} m (dN/dm) dm} 
{\int_{m_1}^{m_2}(dN/dm) dm},
\end{equation} 
where $dN/dm$ is the IMF. For the ``diet Salpeter'' IMF (which, again,
gives results similar to the Kroupa IMF assumed by the VESPA SFH 
reconstruction), and an initial  mass range of 3--8 M$_\odot$, often considered
for the primary stars of SN~Ia progenitor systems, the ratio  of the two
integrals equals 33. Our present result  of $N_{\rm Ia}/M=0.00130\pm 0.00015~{\rm M}_\odot^{-1}$
translates to $\eta=4.3\pm 0.5$\%.
Interestingly, Mannucci et al. (2006) estimated $\eta=4.3\%$ by  
modeling SN Ia rates per unit mass as a function of galaxy colors in a local sample.

\subsection{Analysis of cross-talk between bins}

%\begin{figure}
%\includegraphics[angle=-90, width=0.55\textwidth]{../snloss/sdss2fig2.ps}
%\caption{Recovered DTD for a sample of SDSS2 core-collapse SNe (filled red circles),
%scaled to have a level of 1 in the first time bin. Also shown (empty black circles) 
%is the DTD for CC-SNe for the full LOSS sample (M11), and the CC-SN DTD for the culled
%sample of M11 (empty green squares), which excludes galaxies of type Sab-Sbc. 
%For CC SNe, any nonzero signal of the DTD in 
%the second and third bins is due to ``leakage'' from the prompt bin resulting from 
%incorrect characterization of the full stellar populations of the
% galaxies by the limited fibre aperture of the SDSS spectra. The SDSS2 sample, owing to its much higher redshift, should be less
%susceptible than the LOSS sample to this type of cross talk. This is confirmed in the figure,
%which shows that, in the culled M11 sample,
%the leakage in the SDSS2 sample is lower than in the full M11 sample, and consistent with zero.  
%}
%\label{sdss2dtdcc}
%\end{figure}

In the M11 analysis of the LOSS SN sample and the VESPA SFHs of the monitored LOSS galaxies,
a major systematic error emerged, due to the limited $3''$ aperture size of the SDSS fibres. 
The LOSS galaxies are 
relatively nearby, $z<0.05$, with half of them at $z<0.022$, and are hence large in angle.
The SDSS 
spectra in many cases are therefore
dominated by old populations in the centres of the galaxies probed by the fibres.
However,
many of the LOSS SNe exploded in the outer regions, where there is ongoing star
formation outside the aperture of the SDSS spectroscopy. As a result, the
DTD solution mistakenly associated these SNe with an old population,
and hence a large delay. The result was an artificially enhanced DTD
amplitude in the intermediate and delayed bins, at the expense of
the prompt bin, an effect referred to as DTD inter-bin ``leak'' or ``cross-talk'' in M11.
To mitigate this effect, M11 defined a culled sample of galaxies that excluded Sab-Sbc
Hubble types, which are most susceptible to this stellar-population misrepresentation
by the fibre spectrum. The DTD derived from the culled galaxy sample, and the SNe 
that they hosted, had a much-reduced DTD leak, but at the price of a reduced sample
of SNe, resulting in larger DTD Poisson errors.  

The DTD inter-bin leak, and its reduction in the culled sample, was diagnosed in M11 
by deriving the DTD for core-collapse (CC) SNe.
 Core-collapse SNe explode
within $\lesssim 40$~Myr of star formation. The CC SN DTD should
have zero amplitude on timescales much longer than this, particularly in the bins
at $t>420$~Myr. Thus, any signal in those bins is due to a leak.
Compared to LOSS, we expect the SDSS2 sample to have smaller or no inter-bin
DTD leaks. The typical redshift of the SN hosts is $z=0.15$, an order
of magnitude larger than for LOSS, and all but three of the SDSS2 SN hosts are at $z>0.05$. The
spectra from the SDSS fibres should thus represent well the entire stellar population of each 
galaxy. To verify this, we have compiled a sample of all SNe in the SDSS2 website's list
that have been spectroscopically classified as CC SNe and are associated
with a Stripe 82 galaxy with a VESPA SFH. There are 41 such CC SNe. This list is, of  
course, far from complete, particularly because the main SDSS2 science objective was
the discovery of SNe Ia, and events suspected to be CC SNe had low priority for
spectroscopic followup (D10). Nevertheless, even in a small and incomplete sample,
DTD leak will still manifest itself in the form of a DTD signal at long delays.

The leak effect can be quantified via the ratio between the
intermediate and the prompt
CC-SN DTD components, $\Psi_2/\Psi_1$, which should be zero when there
is no leak of signal from the first to the second DTD bin.
In the full LOSS CC sample of M11 (see their table~1), $\Psi_2/\Psi_1 = 0.165\pm0.045$,
i.e., there is clear presence of a leak. In the M11 subsample that was culled
of Sab-Sbc Hubble types, this ratio was reduced to
$\Psi_2/\Psi_1=0.077\pm 0.050$, which is consistent with zero.
For the SDSS2 CC-SN sample, we find $\Psi_2/\Psi_1=0.069\pm 0.050$,
similar to the ratio in the culled LOSS sample, and again 
consistent with zero.  

As another test of cross-talk, we have recalculated the SN Ia DTD, limiting the
galaxies and the SNe to be above some redshift threshold. Since the fibre-related 
DTD leak decreases with increasing galaxy redshift (as more of the
galaxy's extent 
is included in the aperture),
we expect a change in the DTD when changing thresholds, as was indeed seen in M11.
In the SDSS2 sample, however, we see no significant changes in the
SN~Ia DTD
 when limiting 
the sample to higher redshifts.   
We conclude that the SDSS2 sample's DTD is largely
free of the inter-bin cross talk due to the SDSS fibre aperture.

%Figure~\ref{sdss2dtdcc} shows the SDSS2 DTD for the above CC SN sample. Also
%shown are the CC DTDs from M11 for full LOSS sample, and for the culled M11 sample
%that excludes Sab-Sbc. For a comparison of the leaked fraction,
%all three DTDs have been scaled such that the signal in the prompt bin is 1.
%Clearly, the leak in the SDSS2 galaxies is simlar to that    
%in the culled LOSS sample. In both, the DTD amplitudes in both non-prompt bins are consistent 
%with zero. 
% 0.016\pm0.0038, 0.0011\pm 0.0008, 0.000014\pm 0.000077
% 770 100    127 35   8.6 8.5
% 1320 230    101 66  <5.3

\subsection{Stretch-dependent DTD}
Most DTD recovery 
work to date has considered only 
 a univariate, universal, DTD.
A correlation has been known for some time
(e.g. Hamuy et al. 2000; Neill et al. 2009; Hicken et al. 2009)
 between the age of  a SN Ia host and the  SN
luminosity (or, equivalently, its 
light curve's ``stretch parameter'', $s$, or  
the mass of radioactive Ni synthesized). Old galaxies tend to 
host low-luminosity SNe~Ia, while star-forming galaxies
tend to have luminous SNe Ia. This is an observed link between the
progenitor population and the energy of the explosion, and thus
 may provide a critical clue to the progenitor question.
However, it is only through the DTD recovery approach that one can 
distill this link from the data, decoupled from the SFHs of the individual galaxies. 
B10 used the stretches of their SN Ia sample's
light curves to divide their sample into high-stretch
and low-stretch subsamples, and derived the DTD for
each subsample. 
This was the
first derivation of a 
bivariate $\Psi(t, s)$ distribution, one with 
three delay-time bins and two stretch bins. Here, ``DTD'' is no longer an
appropriate name, as this is now the bivariate distribution (or
  ``response function'') of 
delay times and stretches. 
 The bivariate
response contains information
that is additional to the distribution's
 univariate projection, the DTD, as it gives
not only the age of the progenitor systems but also the run of
explosion energies for each progenitor age. 
B10 indeed found that luminous, high-stretch,
 SNe Ia tend to have most
of their DTD power at short delays, while low-stretch, underluminous,
SNe Ia have a DTD that peaks in the longest-delay bin. 

To test this result with our newly defined SDSS2 sample and analysis, 
we have created low-stretch and high-stretch 
SN Ia subsamples, adopting, like B10, a border between
the two populations of $s=0.92$. Among the 61 of our SNe that are from
the D10 sample with spectroscopic
confirmation, 50 are in the B10 sample, and we adopt the
B10 classifications for these SNe as high- or low-stretch. We exclude the remaining 11
D10 SNe from the present analysis. For the 62 photometrically
classified SNe Ia that are from S11, we convert the $\Delta M_{15}$ 
values, tabulated by S11, to stretch, $s$, using (Guy et al. 2005, their
footnote 7),
\begin{equation}
s=-0.59~\Delta M_{15}+1.55 .
\end{equation}
The 9 additional SNe from our full sample, that were photometrically
classified by D10, do not have stretch, or analogous, parameters
tabulated, so we also exclude them from the present analysis. 
With this selection, we obtain 45 low-stretch SNe Ia 
and 67 high-stretch SNe. 

Table~\ref{table2} lists the DTDs we 
obtain separately for the low-stretch and high-stretch subsamples.
The high-stretch DTD has a $3.5\sigma$ signal in the prompt time
bin, and $\sim 2.5\sigma$ detections in the later time bins.
The best-fitting power law to this DTD has an index $-1.12\pm 0.13$.
The low-stretch DTD has only $1.5\sigma$ detections in the 
prompt and intermediate time bins. Those DTD components are thus 
consistent with zero, but higher $\Psi_1$ and $\Psi_2$ values, 
giving similar component ratios 
as in the high-stretch DTD, also cannot be ruled out. The
delayed low-stretch component, $\Psi_3$, does have a $>3\sigma$
signal. A power-law index in this case is $-0.9^{+0.3}_{-0.15}$.   

Qualitatively, these results are similar to those of B10, in the sense
that the high-stretch SNe have a DTD with a strong prompt signal,
while the low-stretch SNe have a strong signal in the delayed bin.
However, the present analysis does not reproduce the nearly one-to-one correspondence
reported by B10, where essentially all high-stretch SNe derive from 
a young population, and all low-stretch SNe from an old one (see
also further discussion of the B10 results in
Section~\ref{algocompare},
below). 
Rather, in the present analysis, there may be a trend in the above sense, but high-stretch SNe
do have significant non-zero amplitudes in their intermediate and
delayed DTD bins. And, within the current uncertainties, low-stretch
SNe may have non-zero prompt and intermediate-delay DTD components.

We suspect that the difference between the present results and those of 
B10 may be due to selection effects that arise from the use, by B10, of a 
sample that includes only spectroscopically confirmed SNe Ia. 
The SDSS2 survey was strongly oriented toward dicovering SNe Ia 
for cosmographic purposes.
The selection of candidates for spectroscopic followup, and
the success of those followup observations, were both likely tied
to the luminosity of the SN (and hence to its stretch) and to the
type of host galaxy. For example, a candidate in an early-type galaxy would have been
likely considered a promising SN Ia, even if its luminosity was not 
high. Its spectroscopy may have been more likely to succceed
in the presence of the lower host starlight background, compared to a 
similar low-luminosity  candidate in a star-forming galaxy. Such biases
could artificially eliminate, in the spectroscopically confirmed sample,
 low-luminosity SNe Ia in star-forming galaxies, and this would 
reduce or eliminate the $\Psi_1$ component in the low-stretch sample.
The more-complete, spectroscopic plus photometric, sample analyzed
here is more 
immune to such a possible selection effect, and may thus be reflecting 
better the true DTDs. Recently, Perrett et al. (2012) have also failed
to find strong differences in volumetric rate evolution between
low-$s$ and high-$s$ SNe Ia.
Larger, well-defined, samples will be able
to further probe the bivariate distribution of SN Ia delay and stretch. 
 
%{\bf Full}&66,000&132&$140\pm30 $&$25\pm 6$&$1.8\pm0.4$&$0.00130\pm0.00015$\\
%High $s$&66,000&67&$87\pm24 $&$10\pm 4$&$0.8\pm0.3$&$0.00066\pm0.00011$\\
%Low $s$&66,000&45&$24\pm15 $&$6\pm 4$&$1.0\pm 0.3$&$0.00032\pm0.00007$\\

There is some concern that the DTD that we
have recovered may be biased, due to the known correlation between
host galaxy age, or star-formation rate, on the one hand,
and SN stretch or luminosity, on the other. 
The D10 detection efficiency function
that we have used does not account for this correlation. Thus, for
example, early-type galaxies, which tend to host lower-luminosity SNe,
would have a true detection efficiency lower than that of D10. Our
overestimate of the efficiency could then lead to a perceived paucity
of SNe Ia in old populations, and hence to an underestimate of
the true $\Psi_3$ DTD component. To gauge the possible magnitude of such an
effect, we have repeated the DTD derivation, but modifying the D10
detection efficiency function as follows. Sullivan et al. (2006)
have shown (their fig. 11) that galaxies with specific
star-formation rates ${\rm sSFR}>10^{-9.5}{\rm yr}^{-1}$
host SNe with a mean stretch of $s\approx 1.03$, while the most 
passive galaxies, with ${\rm sSFR}<10^{-12}{\rm yr}^{-1}$ have 
SNe with mean $s\approx 0.87$. This change in
$s$ corresponds to a change in SN peak absolute magnitude by about 
0.25 mag. Such a change in brightness, in turn, corresponds to a change in
redshift of $\Delta z/z=\pm 6\%$.
For galaxies that, based on their VESPA SFHs, have current specific
star-formation rates ${\rm sSFR}>10^{-9.5}{\rm yr}^{-1}$, we therefore
modify the D10 efficiency function (described in
Section~\ref{galaxysample}), 
such that it declines linearly 
from $z=0.184$ (rather than $z=0.175$), reaching zero at $z=0.42$
(rather than $z=0.4$). For the passive galaxies
with  ${\rm sSFR}<10^{-12}{\rm yr}^{-1}$, we assume a 
detection function that declines linearly from $z=0.166$, reaching
 zero already at $z=0.38$. For the intermediate, weakly star-forming galaxies,
we use the D10 detection efficiency, as before. 
We find that the recovered DTD is only slightly affected by this
modification of the detection efficiency, with $\Psi_1$ decreasing by 
10\% and $\Psi_3$ increasing by 10\%. 
The change
leads to a slightly less steep power-law fit to the DTD, with an index
of $\beta=-1.07\pm 0.07$.

\subsection{Comparison of the B10 and M11 DTD recovery methods}
\label{algocompare}
As     elaborated above, the SN Ia sample and the details of
the DTD calculation based on SDSS2 SNe differ in a  number of ways between the present work and that 
of B10. Nevertheless, it is interesting to compare the performance and results 
of the two different algorithmic approaches -- on the one hand, that of M11 and of this
work, i.e., comparing observed and predicted (given a DTD model)
SN numbers in each galaxy; and on the other hand that of B10, comparing the observed mean 
host spectrum to a mock mean host spectrum (for a given DTD model). We have therefore calculated
a DTD reconstruction with the M11 approach, but reproducing as closely as possible 
the SN Ia sample and the assumptions of B10. 

To construct a SN Ia sample analogous to the one of B10, we have matched 
coordinates and redshifts between
the 101 SNe Ia in table 2 of B10 and the current Stripe 82 VESPA sample of galaxies 
(without any RA limits). We have matched 89 of the B10 SNe to host
galaxies (we note that 37 of the SNe in B10 have mis-typed declination
signs -- positive instead of negative). This 
smaller number of SNe, 89 compared to 101, is expected, given our
smaller galaxy sample (66,000) 
compared to that of B10 (77,000). Among these 89 SNe Ia, B10 listed 32
as low-stretch and 57 as high-stretch.
We have then run these low-stretch and high-stretch samples through
the M11 DTD recovery algorithm, 
but using 
the same 
efficiency function assumed by B10, with the same parameters that they
assumed 
for high-stretch and low-stretch SNe. In this calculation, we 
do not correct for cosmological time dilation, and do not scale down the VESPA masses
by 0.55 (see \S2, above). We find
for the three of components of the DTD, $\Psi_1$, $\Psi_2$, and $\Psi_3$, respectively, 
in units of $10^{-14}$ yr$^{-1}$M$_\odot ^{-1}$,\\
Low-stretch: ($0\pm 28;~~~   0\pm 5;~~~     1.3\pm0.4$);\\
High-stretch: ($65\pm 15;~~~   8\pm 3;~~~     0.1\pm0.2$).\\
%The results of B10 are given separately for low-stretch and high-stretch SNe.
%To compare to the total DTD, we add their two stretch components for each 
%time bin, and add the uncertainties in quadrature. 
Table 3 and figure 7 of B10 give, for several of the  low-stretch and high-stretch  
{\bf $\Psi$} components ($\epsilon_{h,2}$, and $\epsilon_{l,2}$, in their
nomenclature) their results and uncertainties. To these uncertainties we have added
$1\sigma$ Poisson errors, not included in B10. 
The B10 $\Psi_i$ results are,\\
%              0\pm3     0\pm2      1.6\m0.1
Low-stretch: ($0\pm20;~~~ 0\pm 5;~~~ 1.6\pm0.3$);\\
High-stretch: ($75\pm19;~~~ 8\pm 5;~~~ 0.3\pm0.2$).\\
%                           8\pm 5     0.3\pm0.2
We see that the best-fit results and the uncertainties found by
 the two algorithms are quite similar. The slightly higher values 
found by B10 for the significantly non-zero components
 may be due to the fact that B10 draw the DTD
values that they test from an assumed prior distribution, while in M11
and here we
do an adaptive grid
search for the best fitting DTD in the 3D space of $\Psi_1$, $\Psi_2$,
and $\Psi_3$.        

\section{Discussion and summary} 
We have taken the DTD recovery method presented in M11, where it
 was applied to
the LOSS SN sample, and applied it here to a subsample of SDSS2 SNe Ia that were hosted by
Stripe-82 SDSS galaxies with VESPA SFH reconstructions. Among
all of the DTD recontructions to date that are  based on detailed SFHs of individual
galaxies (B10; M11; Maoz \& Badenes 2010; this work), we have obtained 
the DTD that is the most precise (in terms of Poisson errors) and the most
accurate (in terms of control of systematic errors).
The full M11 sample was systematically affected by the limited SDSS fibre
aperture, which led to cross talk between DTD bins. In the reduced M11
sample, where this problem was mitigated, only 49 SNe Ia remained,
leading to large Poisson errors. In the analysis by Maoz \& Badenes
(2010) of 77 Magellanic Cloud SN remnants, the bulk of the DTD signal
was at $t<35$~Myr, corresponding to core-collapse SNe. This meant that 
only of order 10 SNe Ia were effectively driving the SN Ia DTD, which 
was therefore very noisy. Another possible systematic error in Maoz \& Badenes
(2010) was contamination of the SNa Ia DTD by CC SNe, with progenitors
near the poorly known progenitor age border between CC-SNe  and WD formation.

The SNe from SDSS2, which are at larger distances and are well classified,
 can avoid the above systematics, and at the same time 
allow for relatively large numbers of SNe Ia. The B10 analysis of SDSS2 SNe, however,  
used a limited sample,  available
at the time, consisting only of spectrally confirmed SNe. 
Furthermore, B10 estimated a detection efficiency function that did
not take into account the details of the real observing patterns,
the detection process, classification, and sample selection, details that were later
quantified by D10, based on real-time detection simulations. Finally, the B10
analysis omitted  necessary corrections to the SN
visibility times 
and to the stellar masses of the galaxies. All of the above are
potential sources of systematic error, affecting each DTD component
by up to a factor $\sim 2$. In our renewed definition and 
analysis of an SDSS2 SN sample, we have addressed these issues. 
In this analysis, we have
detected for the first time, at a highly significant level, 
 an intermediate-delay DTD component, at $0.42<t<2.4$~Gyr. 
B10 and M11
observed such a component only at the $\sim 2\sigma$ level.
In addition, the completeness of the sample and the 
consideration of  absolute stellar mass, visibility time, and
efficiency, permit a robust derivation of absolute DTD levels. 

It is interesting to consider the cause behind the much more significant detection
of an intermediate-delay DTD component, $\Psi_2$, compared to the B10
analysis of a differently defined SDSS2 SN Ia sample. Given that the M11 and B10 algorithms
give similar results and uncertainties (Section~\ref{algocompare},
above), the difference in the $\Psi_2$
significance could be due to
the differences in the SN samples, the differences in the adopted
efficiency functions, or both. To investigate this, we have recovered
best-fit 
DTDs with the present SN Ia sample (Table~1), but using the B10 efficiency function, 
and omitting other corrections, as in
Section~\ref{algocompare}. Alternatively, we have used the B10 SN sample
as described in Section~\ref{algocompare}, but using the empirical D10
efficiency function, $1+z$ correction, and so on, as in our main
calculation. We find that neither choice, alone, raises much the 
$\Psi_2$ detection significance. Apparently, the improvement results
from  the combination 
of a carefully selected, relatively complete, sample of SNe, and 
a realistic efficiency function, one  that embodies the selection effects
of that same SN sample. The two, together, sharpen the correlation 
between the presence of SNe Ia and an intermediate-age
stellar population in some galaxies, and thus reveal clearly the $\Psi_2$ component.

Our findings point to a continuous
distribution of SN Ia delay times. Although recent observational
developments have led to discussions of ``prompt and delayed SNe Ia'',
``two SN Ia channels/populations'', ``a bimodal DTD'', and so on, 
 theoretical binary population synthesis
models have almost always predicted a broad and continuous DTD, even
for a single physical channel (e.g. DD). Thus, the continuous
observational DTDs that are emerging (Totani et al. 2008; Maoz et
al. 2010; this work) are not unexpected. Naturally, DTDs recovered 
in the future with finer
time resolution may still reveal a real bimodality. The current coarse
time resolution of the DTD is dictated by the resolution of the galaxy SFHs.
Future improved data will also permit recovery of a better-resolved,
and higher signal-to-noise, bivariate distribution of delay and
stretch, $\Psi(t,s)$, that could sharpen the current hints of a
relation between the progenitor age and the explosion energy.

As in most of the recent studies, the DTD we find appears to be
consistent with the $\sim t^{-1}$ form generally expected in the DD
scenario. This result joins others that are strongly suggestive of
DD progenitors for at least some SNe~Ia. Most notable among these are the absence of 
any signs of the putative donor star, a few years before, and during the
explosion of SN2011fe (Li et al. 2011c, Nugent et al. 2011, Horesh et al. 2012), 
and 400 years after the explosion that left
the SN remnant 0509$-$67.5 (Schaefer \& Pagnotta 2012). 
It is noteworthy that the main objection to
the DD model has been theoretical, arguing that a DD merger cannot 
produce an event with the characteristics of SNe Ia. The fact that 
there are, by now, a couple of cases where we may have witnessed DD mergers
producing normal SNe Ia suggests that the theoretical objection could be
somehow flawed (unless the absence of a companion, in these cases, is the result  
of one of the ``SN on hold'' scenarios mentioned in Section~1, assuming they 
work). If indeed the DD scenario is physically possible after all,
perhaps most or even all SNe Ia could
result from  DD mergers. Examining the evidence from the opposite 
direction, Maoz et al. (2012) and Badenes \& Maoz (2012)
have recently estimated the specific merger rate of Galactic WD binaries.
They found that, if one considers all WD mergers, most of which have
sub-Chandrasekhar merged masses, but only a few tenths of a solar mass below
this limit, their merger rate 
is remarkably similar to the specific SN Ia rate
of Milky-Way-like galaxies. Naturally, this coincidence is meaningful
only if sub-Chandrasekhar CO-CO WD mergers 
can indeed produce a SN Ia, which is
still an open question (e.g. van Kerkwijk et al. 2010; Guillochon et
al. 2010; Pakmor et al. 2011).       

Another interesting result of our analysis is the absolute DTD
normalization that we find -- lower than most of the recent determinations based
on SN Ia rates in nearby galaxies and in galaxy clusters, but consistent with the
levels found based on volumetric SN Ia rates. This result relaxes some of  the tension between the
previous, high,
observational SN Ia production efficiencies, and the lower predictions from binary
population synthesis models (see, e.g. Maoz 2008; Mennekens et al. 2010). 
On the other hand,
our result re-floats the possibility that SNe Ia, for some reason, are produced more
efficiently in galaxy cluster environments than in field environments. 
If this is true, it could be the result of an IMF that depends on
environment. Evidence has accumulated recently that the most massive
elliptical galaxies have ``bottom-heavy'' IMFs, with low-mass
components as steep, or even steeper than, the Salpeter IMF (e.g. Treu et
al. 2010; Cappellari et al. 2012; Conroy \& van Dokkum 2012). Clusters
host the most massive ellipticals. Thompson (2011) has proposed that 
a large fraction of SN Ia progenitor systems are triples, in which a
low-mass tertiary induces ellipticity in the inner WD binary via the
Kozai (1962) mechanism, significantly increasing the time-integrated
number of DD systems that merge (and presumably produce SNe Ia). A speculative
possibility that we raise is that, in cluster galaxies, the large availability
of low-mass stars translates into more tertiaries, and hence a larger
SN Ia production efficiency. 
%On the other hand, this idea is at odds
%with the result of Li et al. (2011b) that, the more massive ellipticals
%are, the lower their SN Ia rates per unit stellar mass.   
   
To summarize, we have constructed a new SDSS2 SN Ia sample that is more
complete and has a better-characterized detection efficiency than what
was available in 
B10. We have analyzed it with the methodology of M11, but avoiding the 
small sample size and the systematic errors in SFH that were
unavoidable in the LOSS sample of M11, 
while making several necessary corrections that
were omitted in B10.
In the end, we find a SN Ia DTD that is more accurate and more precise
than the ones found by either B10 and M11. For the first time, a DTD
based on the three time bins used here shows a highly significant signal 
in all three bins. The DTD is consistent in form with 
the $t^{-1}$ found in the studies reviewed above (including B10 and
M11), but differs in absolute normalization from some of those
studies.

The SN Ia progenitor problem is far from
solved. However, our results demonstrate how our view of the SN Ia DTD
can progressively improve by means of analysis of samples of SNe 
that are progressively larger, purer, more complete, and well-characterized,
and with samples of monitored galaxies with SFHs that are better
determined, systematically, temporally, and spatially. These improved
views will continue to play a role in resolving the progenitor issue. 

\section*{Acknowledgments}
We thank Rita Tojeiro and Keren Sharon for their valuable input to
this work, and the referee for useful comments.
 D.M acknowledges support 
by a grant from the Israel Science Foundation. The work of T.D.B. is
 supported
by a National Science Foundation Graduate Research Fellowship under
 Grant No. DGE-0646086. 
  Funding for the SDSS and SDSS-II has been provided by the Alfred P. Sloan Foundation, the Participating Institutions,
  the National Science Foundation, the U.S. Department of Energy, the National Aeronautics and Space Administration, the
  Japanese Monbukagakusho, the Max Planck Society, and the Higher Education Funding Council for England. The SDSS Web
  Site is http://www.sdss.org/.
  The SDSS is managed by the Astrophysical Research Consortium for the Participating Institutions. The Participating
  Institutions are the American Museum of Natural History, Astrophysical Institute Potsdam, University of Basel,
  University of Cambridge, Case Western Reserve University, University of Chicago, Drexel University, Fermilab, the
  Institute for Advanced Study, the Japan Participation Group, Johns Hopkins University, the Joint Institute for Nuclear
  Astrophysics, the Kavli Institute for Particle Astrophysics and Cosmology, the Korean Scientist Group, the Chinese
  Academy of Sciences (LAMOST), Los Alamos National Laboratory, the Max-Planck-Institute for Astronomy (MPIA), the
  Max-Planck-Institute for Astrophysics (MPA), New Mexico State University, Ohio State University, University of
  Pittsburgh, University of Portsmouth, Princeton University, the United States Naval Observatory, and the University of
  Washington.

%\onecolumn
%\newpage

\end{document}